\documentclass[useAMS,usenatbib]{mn2e}
\pdfoutput=1
\usepackage{graphicx}
\usepackage{journals}
\usepackage[flushleft]{threeparttable}
\usepackage{color}

\def\5{\footnotesize V\normalsize}
\def\4{\footnotesize IV\normalsize}
\def\3{\footnotesize III\normalsize}
\def\2{\footnotesize II\normalsize}
\def\1{\footnotesize I\normalsize}

\def\kms{$\mbox{km s}^{-1}$}

\def\pp{$\phantom{-}$}
\def\o{$\phantom{0}$}

\title[Chemistry and Kinematics of NGC\,2100]{Chemistry and Kinematics of Red Supergiant Stars in the Young Massive Cluster NGC\,2100}
\author[L.~R.~Patrick et al.]{L.~R.~Patrick$^{1}$\thanks{E-mail: lrp@roe.ac.uk},
C.~J.~Evans$^{1, 2}$,
B.~Davies$^{3}$,
R-P.~Kudritzki$^{4,5}$,
V.~H{\'e}nault-Brunet$^{6}$,
\newauthor N.~Bastian$^{3}$,
E.~Lapenna$^{7,8}$, M.~Bergemann$^{9}$,\\
$^{1}$Institute for Astronomy, University of Edinburgh, Royal Observatory Edinburgh, Blackford Hill, Edinburgh EH9 3HJ, UK\\
$^{2}$UK Astronomy Technology Centre, Royal Observatory Edinburgh, Blackford Hill, Edinburgh EH9 3HJ, UK\\
$^{3}$Astrophysics Research Institute, Liverpool John Moores University, Liverpool Science Park ic2, 146 Brownlow Hill, Liverpool L3 5RF, UK\\
$^{4}$Institute for Astronomy, University of Hawaii, 2680 Woodlawn Drive, Honolulu, HI, 96822, USA\\
$^{5}$University Observatory Munich, Scheinerstr. 1, D-81679, Munich, Germany\\
$^{6}$Department of Physics, Faculty of Engineering and Physical Sciences, University of Surrey, Guildford, GU2 7XH, UK\\
$^{7}$Dipartimento di Fisica e Astronomia, Universit\'a degli Studi di Bologna, Viale Berti Pichat 6/2, I-40127 Bologna, Italy\\
$^{8}$INAF-Osservatorio Astronomico di Bologna, via Ranzani 1, I-40127, Bologna, Italy\\
$^{9}$Max-Planck Institute for Astronomy, D-69117, Heidelberg, Germany\\
}

\begin{document}

\date{Accepted  Received 1; in original form}

\pagerange{\pageref{firstpage}--\pageref{lastpage}} \pubyear{2016}

\maketitle

\label{firstpage}

\begin{abstract}
\noindent We have obtained {\it K}-band Multi-Object Spectrograph (KMOS) near-IR spectroscopy for 14 red supergiant stars (RSGs) in the young massive star cluster NGC\,2100 in the Large Magellanic Cloud (LMC).
Stellar parameters including metallicity are estimated using the {\it J}-band analysis technique, which has been rigorously tested in the Local Universe.
We find an average metallicity for NGC\,2100 of [Z]~=~$-$0.43\,$\pm$\,0.10\,dex, in good agreement with estimates from the literature for the LMC.
Comparing our results in NGC\,2100 with those for a Galactic cluster (at Solar-like metallicity) with a similar mass and age we find no significant difference in the location of RSGs in the Hertzsprung--Russell diagram.
We combine the observed KMOS spectra to form a simulated integrated-light cluster spectrum and show that, by analysing this spectrum as a single RSG, the results are consistent with the average properties of the cluster.
Radial velocities are estimated for the targets and the dynamical properties are estimated for the first time within this cluster.
The data are consistent with a flat velocity dispersion profile, and with an upper limit of 3.9\,\kms, at the 95\% confidence level, for the velocity dispersion of the cluster.
However, the intrinsic velocity dispersion is unresolved and could, therefore, be significantly smaller than the upper limit reported here.
An upper limit on the dynamical mass of the cluster is derived as
$M_{dyn}$~$\le$~$15.2\times10^{4}M_{\odot}$ assuming virial equilibrium.
\end{abstract}

\begin{keywords}
stars: abundance, (stars:) supergiants, (galaxies:) Magellanic Clouds, galaxies: star clusters: individual: NGC\,2100
\end{keywords}

\section{Introduction} 
\label{sec:introduction}

Young massive clusters (YMCs\footnotemark) are important probes of the early evolution of star clusters and have increasingly been used as tracers of star formation in galaxies~\citep[e.g.][]{1995AJ....109..960W,1997AJ....114.2381M,1999AJ....118..752Z}.
Known to contain large populations of massive stars, YMCs are also important tracers of massive star formation, which is heavily clustered~\citep{2003ARA&A..41...57L,2005A&A...437..247D,2007MNRAS.380.1271P}.
In addition to being the birthplace of most of the massive stars in the Local Universe~\citep[$>200\,$M$_{\odot}$ stars in R136;][]{2010MNRAS.408..731C}, owing to the density of stars, YMCs are thought to be the birthplace of some of the rich stellar exotica
(e.g. blue stragglers, X-ray binaries and radio pulsars) found in the old population of globular clusters~\citep[GCs;][]{2010ARA&A..48..431P}.

\footnotetext{A YMC is defined as having an age of $<100\,$Myr and a stellar mass of $>10^{4}\,$M$_{\odot}$~\citep{2010ARA&A..48..431P}.}

Recently, the idea that GCs are simple stellar populations has been called into question based on chemical anomalies of light elements~\citep[C, N, O, Na and Al; e.g.][]{2012A&ARv..20...50G}.
These anomalies are considered by most authors to be the signature of multiple stellar populations within GCs.
Studying YMCs could therefore potentially help to constrain some of the proposed models for creating multiple stellar populations within GCs~\citep[e.g.][]{2014MNRAS.441.2754C}.

Investigating the link between YMCs and older clusters is an important, uncertain, factor in the evolution of young clusters.
As most stellar systems are thought to dissolve shortly after formation
\citep{2003ARA&A..41...57L}, determining how long bound systems can remain so is an important question to answer.
Studying the dynamical properties of YMCs is, therefore, an important tool to evaluate the likelihood that young clusters will survive.
In addition, the study of YMCs in different environments can help bridge the gap between the understanding of star formation in the Solar neighbourhood and that in the high-redshift Universe.

Over the last few years, medium resolution ($R~\geq~3000$) near-IR spectroscopy has been shown to be a powerful tool to estimate stellar parameters for red supergiant stars~\citep[RSGs;][]{2010MNRAS.407.1203D}.
RSGs are the final evolutionary stage of a massive star and, owing to their cool atmospheres~\citep[T$_{\rm eff}\sim$~4000\,K;][]{2013ApJ...767....3D}, are brightest at $\sim1.1\,\mu$m.
In star-forming galaxies, RSGs are the most luminous near-IR sources, therefore, they can be observed out to large distances at these wavelengths.
Given that dust extinction is intrinsically lower at near-IR wavelengths and that the next generation of ground-/space-based telescopes will be optimised for observations at these wavelengths, RSGs are likely to become increasingly attractive targets by which to study distant star-forming galaxies.

The $J$-band analysis technique for estimating metallicities and stellar parameters of RSGs has been rigorously tested by~\cite{2014ApJ...788...58G} and~\cite{2015ApJ...806...21D}.
These authors show that metallicities can be estimated in extragalactic systems to a high level of accuracy and to a precision of $<0.15$\,dex.

The availability of the $K$-band multi-object spectrograph~\citep[KMOS;][]{2013Msngr.151...21S} at the Very Large Telescope (VLT), has presented new opportunities for efficient observations of samples of RSGs in external galaxies to study their distribution and build-up of metals.
\cite{2015ApJ...803...14P} used KMOS observations to investigate the present-day metallicity of NGC\,6822 ($d$~=~0.5\,Mpc) and~\cite{2015ApJ...805..182G} determined the metallicity gradient of NGC\,300, a grand design spiral galaxy outside the Local Group ($d$~=~1.9\,Mpc), finding striking agreement with previous measurements from stars and H\,\2 regions.

In addition,~\citet{2013MNRAS.430L..35G} demonstrated that, after the appearance of the first RSGs within a YMC, the overall near-IR flux from the cluster is dominated by the RSGs (F$_{J, RSG}/$F$_{J}>0.90$).
Using this result, these authors showed that the spectrum from an unresolved star cluster can be used to estimate the average properties of the RSG population of the cluster using exactly the same analysis method as for single stars.
\citet{2015ApJ...812..160L} demonstrated this with KMOS spectroscopy of three unresolved YMCs in NGC\,4038 in the Antennae ($d$~=~20\,Mpc), at Solar-like metallicity, finding good agreement with previous studies.
With a multi-object spectrograph operating on the European Extremely Large Telescope, this technique could be used to measure metallicities of individual RSGs at distances of $>10\,$Mpc and from YMCs out to potentially $>100\,$Mpc~\citep{2011A&A...527A..50E}.

\begin{figure}
 \includegraphics[width=9.0cm]{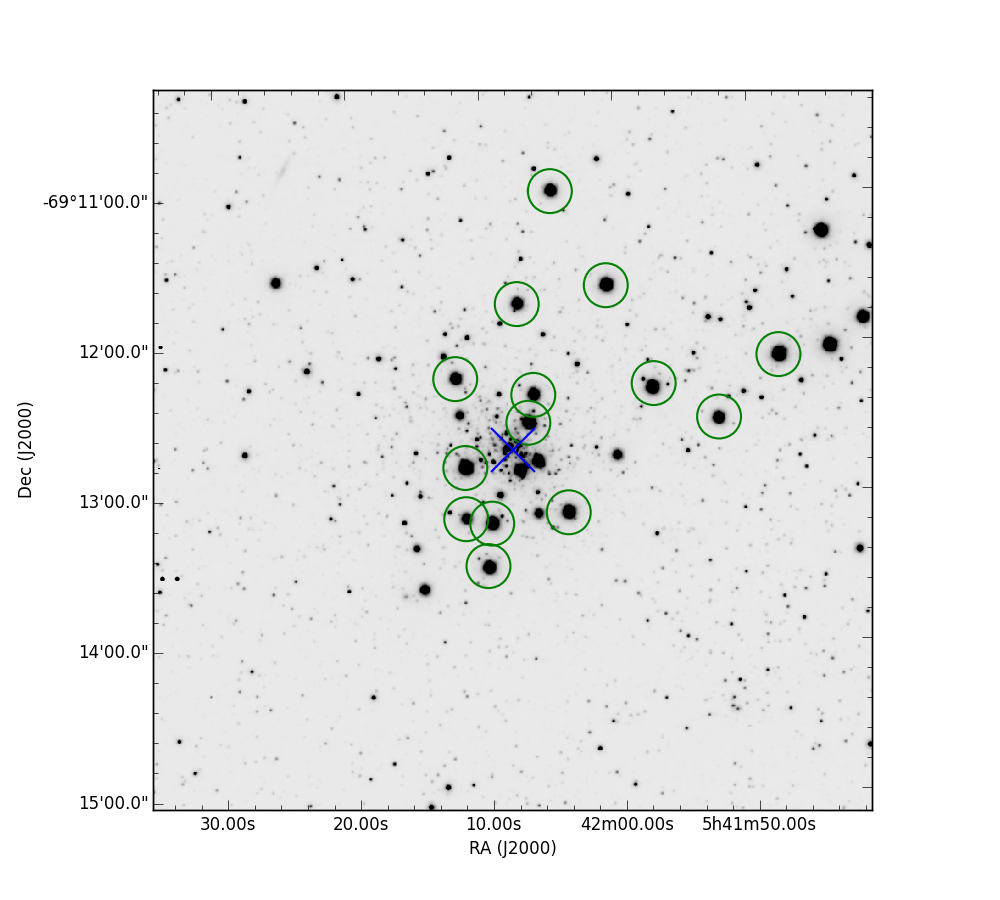}
 \caption{Positions of the KMOS targets in NGC\,2100 overlaid on a VISTA $J$-band image~\citep{2011A&A...527A.116C}.
          Green circles indicate KMOS targets.
          The adopted cluster centre has been marked by a blue cross.\label{fig:targets}
          }
\end{figure}

NGC\,2100 is a YMC in the Large Magellanic Cloud (LMC), located near the large star-forming 30 Doradus region.
With an age of $\sim$\,20\,Myr~\citep{1991ApJS...76..185E,2015A&A...575A..62N}, and a photometric mass of $4.6~\times~10^4M_{\odot}$~\citep[assuming~\cite{1966AJ.....71...64K} profiles]{2005ApJS..161..304M}, NGC\,2100 falls within the mass and age range where the near-IR cluster light is dominated by RSGs~\citep{2013MNRAS.430L..35G}.
This is supported by the large number of RSGs identified within this cluster (see Figure~\ref{fig:targets}).

NGC\,2100 is not a cluster in isolation.
It is located in one of the most actively star-forming regions within the Local Group of galaxies.
At $\sim$\,20\,Myr old, the most massive members of this star cluster will have already exploded as supernovae.
This should have had a profound effect on the surrounding gas and dust, and has potentially shaped the surrounding LMC\,2 supershell~\citep[see][]{1999ApJ...518..298P}.

In this study we estimate stellar parameters from KMOS spectroscopy for 14 RSGs which appear to be associated with NGC\,2100.
Section~\ref{sec:observations} describes the observations and data reduction, and in Section~\ref{sec:results} we detail our results, focusing on radial velocities of the target stars where we derive the line-of-sight velocity dispersion,
the dynamical mass of NGC\,2100 and stellar parameters.
Our results are discussed in Section~\ref{sec:discussion} and conclusions are presented in Section~\ref{sec:conclusions}.


\section{Observations and Data Reduction} 
\label{sec:observations}
These observations were obtained as part of the KMOS Guaranteed Time Observations (PI: Evans 095.B-0022) in March 2015.
The observations consisted of $8\times10$\,s exposures (seeing conditions $\sim$1\farcs0) taken with the $YJ$ grating with sky offset exposures (S) interleaved between the object exposures (O) in an O,~S,~O observing pattern.
In addition, a standard set of KMOS calibration frames were obtained as well as observations of HD\,51506 (B5) as the telluric standard star.
Figure~\ref{fig:targets} shows the observed RSGs overlaid on a {\it J}-band VISTA image of the surrounding region~\citep{2011A&A...527A.116C}.

The standard KMOS/esorex routines~\citep[SPARK;][]{2013A&A...558A..56D} were used to calibrate and reconstruct the data cubes.
Telluric correction was performed using observations of the standard star in all 24 IFUs using the methodology described in detail by
\citet{2015ApJ...803...14P}.
Briefly, corrections are made to the standard telluric recipe to account for slight differences in wavelength calibration between the telluric and science spectra.
This is implemented using an iterative cross-correlation approach.
Additionally, differences in the strength of the telluric features are corrected by applying a simple scaling using the equation:

\begin{equation}
  T_{2} = (T_{1} + c) / (1 + c)
\end{equation}

\noindent where $T_{2}$ is the scaled telluric-standard spectrum, $T_{1}$ is the uncorrected telluric-standard spectrum and {\it c} is the scaling parameter which is varied from {\it c}~=~$-$0.5 to {\it c}~=~0.5 in increments of 0.02.
The best value of {\it c} is chosen based on the overall standard deviation of the spectrum, i.e. the {\it c} value producing the smallest $\sigma$ is selected.
Once these corrections are accounted for, the science spectra are divided by the appropriate telluric spectrum for that particular KMOS integral field unit (IFU).

\begin{table*}
\caption{
        Observed properties of VLT-KMOS targets in NGC\,2100.\label{tb:obs-params}
        }
\scriptsize
\begin{center}
\begin{threeparttable}
\begin{tabular}{lrccccl }
 \hline
 \hline
ID & S/N & $J$\tnote{a} & $H$\tnote{a} & $K_{\rm s}$\tnote{a} & RV (\kms) & Notes\tnote{b} \\
 \hline
J054147.86$-$691205.9 & 320 &\o9.525 &\o8.603 & 8.200 &  250.3 $\pm$ 4.7 & D15\\
J054152.51$-$691230.8 & 200 & 10.413 &\o9.526 & 9.155 &  249.3 $\pm$ 2.6 & D16\\
J054157.44$-$691218.1 & 200 &\o9.811 &\o9.036 & 8.738 &  245.6 $\pm$ 3.5 & C2\\ 
J054200.74$-$691137.0 & 260 &\o9.900 &\o9.017 & 8.683 &  248.8 $\pm$ 2.7 & C8\\
J054203.90$-$691307.4 & 250 &\o9.839 &\o8.996 & 8.740 &  251.1 $\pm$ 2.8 & B4\\
J054204.78$-$691058.8 & 210 & 10.319 &\o9.427 & 9.159 &  256.1 $\pm$ 4.0 & \ldots\\
J054206.36$-$691220.2 & 200 & 10.371 &\o9.480 & 9.159 &  255.7 $\pm$ 4.9 & B17\\
J054206.77$-$691231.1 & 250 &\o9.977 &\o9.150 & 8.807 &  250.6 $\pm$ 3.4 & A127\\
J054207.45$-$691143.8 & 200 & 10.482 &\o9.610 & 9.351 &  252.5 $\pm$ 3.0 & C12\\
J054209.66$-$691311.2 & 240 &\o9.976 &\o9.136 & 8.841 &  254.3 $\pm$ 4.1 & B47\\
J054209.98$-$691328.8 & 250 & 10.021 &\o9.150 & 8.823 &  250.2 $\pm$ 3.0 & C32\\
J054211.56$-$691248.7 & 300 &\o9.557 &\o8.617 & 8.264 &  255.5 $\pm$ 4.3 & B40\\
J054211.61$-$691309.2 & 150 & 10.943 & 10.090 & 9.788 &  256.6 $\pm$ 6.1 & B46\\
J054212.20$-$691213.3 & 200 & 10.440 &\o9.622 & 9.335 &  260.0 $\pm$ 4.8 & B22\\

\hline
\end{tabular}
\begin{tablenotes}
\item [a] Photometric data from 2MASS, with tyipcal errors on $J$, $H$, and $K_{\rm s}$ of 0.024, 0.026 and 0.022\,mag respectively.
\item [b] Cross-identifications in final column from~\cite{1974A&AS...15..261R}.
\end{tablenotes}
\end{threeparttable}
\end{center}
\end{table*}


\section{Results} 
\label{sec:results}


\subsection{Radial velocities and velocity dispersion} 
\label{sub:radial_velocities}
Radial velocities are estimated using an iterative cross-correlation method.
To ensure systematic shifts are removed, the observed spectra are first cross-correlated against a spectrum of the Earth's atmosphere, taken from the European Southern Observatory web pages\footnotemark, at a much higher spectral resolution than that of KMOS.
This spectrum is then degraded to the resolution of the observations using a simple Gaussian filter.
The cross-correlation is performed within the 1.140--1.155\,$\mu$m region, as a strong set of reliable telluric features dominates this region, with minimal contamination from stellar features.
The shift arising from this comparison is typically 0--10\,\kms~and is then applied to the science spectra so that they are on a consistent wavelength solution.

\footnotetext{Retrieved from http://www.eso.org/sci/facilities/paranal/\\
decommissioned/isaac/tools/spectroscopic\_standards.html}

Stellar radial velocities are estimated following a similar approach to the methods used by~\citet{2015ApJ...798...23L} and~\citet{2015ApJ...803...14P}. An initial radial-velocity estimate is found for each star from cross-correlation of the KMOS spectra with an appropriate model spectrum in the 1.16--1.22\,$\mu$m region
(selected owing to the dominance of atomic features in RSG spectra at these wavelengths).
We improved on this initial estimate via independent cross-correlation of the observed and model spectra for seven strong absorption lines in this region.

The quoted radial velocity for each star is the mean of these estimates, where the quoted uncertainty is the standard error of the mean
(i.e. $\sigma$/$\sqrt{n_{\rm lines}}$).
Obvious outliers (with $\delta$RVs of tens of \kms) were excluded in calculating the mean estimates; such outliers arise occasionally from spurious peaks in the cross-correlation functions from noise/systematics in the spectra.

In order to sample from the posterior probability distribution for the intrinsic velocity dispersion and mean cluster velocity (given the observed radial velocity estimates and their uncertainties), we use \texttt{emcee}~\citep{2013PASP..125..306F},
an implementation of the affine-invariant ensemble sampler for Markov chain Monte Carlo (MCMC) of \cite{2010CAMCS.5..65G}. Our likelihood function is given by

\begin{equation}
p(D|\{\sigma_{1D}, v_0\}) = \prod_i \frac{1}{\sqrt{2 \pi (\sigma_{1D}^2+ \sigma_{v, i}^2)}}  \exp{\left(\frac{-(v_i - v_0)^2}{2 (\sigma_{1D}^2+ \sigma_{v, i}^2)}\right)},
\label{eq:like}
\end{equation}

\noindent where $\sigma_{1D}$ is the intrinsic velocity dispersion of the cluster, $v_0$ is the mean cluster velocity, and the data consists of our set of radial velocity measurements $v_i$ and their uncertainties $\sigma_{v, i}$. We therefore assume that the intrinsic cluster dispersion is Gaussian with no variations in the dispersion across the sample.
The systemic radial velocity ($v_0$) of the sample is estimated to be 251.6\,$\pm$\,1.1\,\kms.

\begin{figure}
 \includegraphics[width=9.0cm]{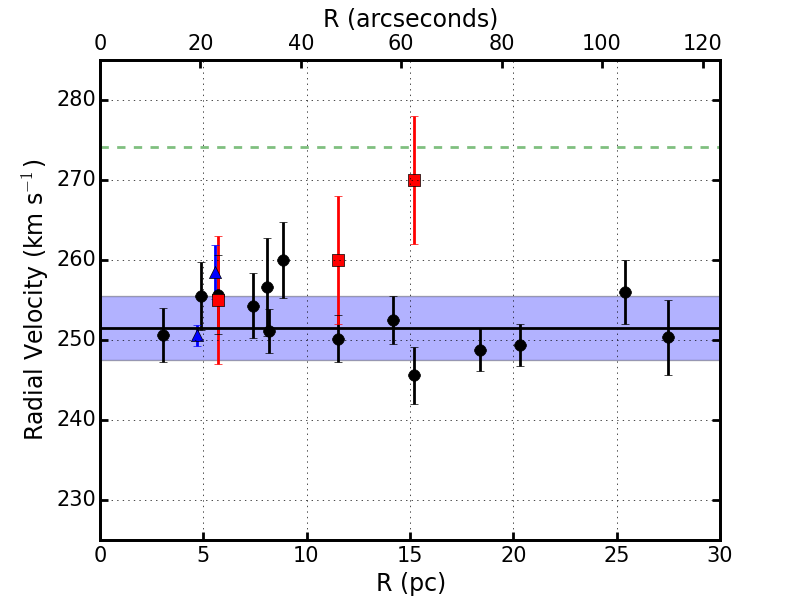}
 \caption{Radial velocities of KMOS targets (black points) shown as a function of distance from the cluster centre.
The green dashed line shows the LMC systemic velocity of $\sim$200 massive stars from
 {\protect\citep[274.1\,$\pm$\,16.4\,\kms;][]{2015A&A...584A...5E}}.
 The solid black line shows the mean cluster velocity ($v_0$~=~251.6\,$\pm$\,1.1\,\kms) and the shaded blue region shows $v_0\,\pm\sigma_{1D}$.
 The blue triangles show estimates for two OB-type stars in NGC\,2100~\protect\citep{2015A&A...584A...5E} and the red squares show previous estimates for three of our targets
 {\citep{1994A&A...282..717J}}.
 The distance modulus used to produce this figure is 18.5~\citep{2013Natur.495...76P,2014AJ....147..122D}.
 \label{fig:rvs}}
\end{figure}

Table~\ref{tb:obs-params} displays our stellar radial-velocity estimates and Figure~\ref{fig:rvs} shows these estimates as a function of distance from the centre of the cluster, compared with the average radial velocity of $\sim$200 massive stars within the LMC from~\citet[][green dashed line]{2015A&A...584A...5E}.
To quantify the likelihood that the measured velocities are consistent with the NGC\,2100 mean cluster velocity we calculate the probability that each measured velocity is drawn from a two-component mixture of Gaussian distributions with
$P(x|\{\mu, \sigma\}_{NGC\,2100}) + P(x|\{\mu, \sigma\}_{LMC-field}) = 1$, where the {\it LMC-field} distribution is defined by~\cite{2015A&A...584A...5E}.

From this analysis one target (J054212.20$-$691213.3) has a measured velocity with greater probability of being drawn from the underlying distribution of massive stars rather than the distribution centred on the NGC\,2100 systemic velocity.
Excluding this target from the sample does not alter the estimation of $v_0$ or $\sigma_{1D}$ significantly, therefore we choose to include this target for further analysis.

We conclude that all targets have a velocity consistent with membership to the LMC (as opposed to Galactic objects) and that none display compelling evidence for being excluded from membership of NGC\,2100.

The estimated $v_0$ is in reasonable agreement with previous measurements for two OB-type stars in the cluster
\citep{2015A&A...584A...5E} as well as the results from four RSGs in NGC\,2100
\citep[henceforth JT94; three of which were observed in the current study]{1994A&A...282..717J}.
Table~\ref{tb:rvs} contains the details of previous radial velocity measurements within NGC\,2100.
We conclude that there exists no significant difference between our measurements and previous estimates within NGC\,2100.
This is an additional confirmation that absolute radial velocities can be precisely measured with KMOS spectra.

\begin{table*}
\begin{center}
\caption{
        Literature stellar radial-velocity measurements within NGC\,2100.\label{tb:rvs}
        }
\scriptsize
\begin{threeparttable}
\begin{tabular}{lcccll}
 \hline
 \hline
\multicolumn{2}{c}{ID} & \multicolumn{2}{c}{RV (\kms)}  & Reference & Notes \\
Lit. & current study & Lit. & current study\\
 \hline
AA$\Omega$\,30\,Dor\,407 & ---         & $258.5\pm3.4$     & \ldots        & {\cite{2015A&A...584A...5E}} &  O9.5\,II  \\
AA$\Omega$\,30\,Dor\,408 & ---         & $250.6\pm1.3$     & \ldots        & {\cite{2015A&A...584A...5E}} &  B3\,Ia    \\
R74\,B17 & J054206.36-691220.2 & $255\pm8$ & $255.7\pm4.9$ & {\cite{1994A&A...282..717J}} \\
R74\,C2  & J054157.44-691218.1 & $270\pm8$ & $245.6\pm3.5$ & {\cite{1994A&A...282..717J}} \\
R74\,C32 & J054209.98-691328.8 & $260\pm8$ & $250.2\pm3.0$ & {\cite{1994A&A...282..717J}} \\
R74\,C34 & ---         & $265\pm8$         & \ldots        & {\cite{1994A&A...282..717J}} & \\

\hline
\end{tabular}
\end{threeparttable}
\end{center}
\end{table*}

\begin{figure}
 \includegraphics[width=9.0cm]{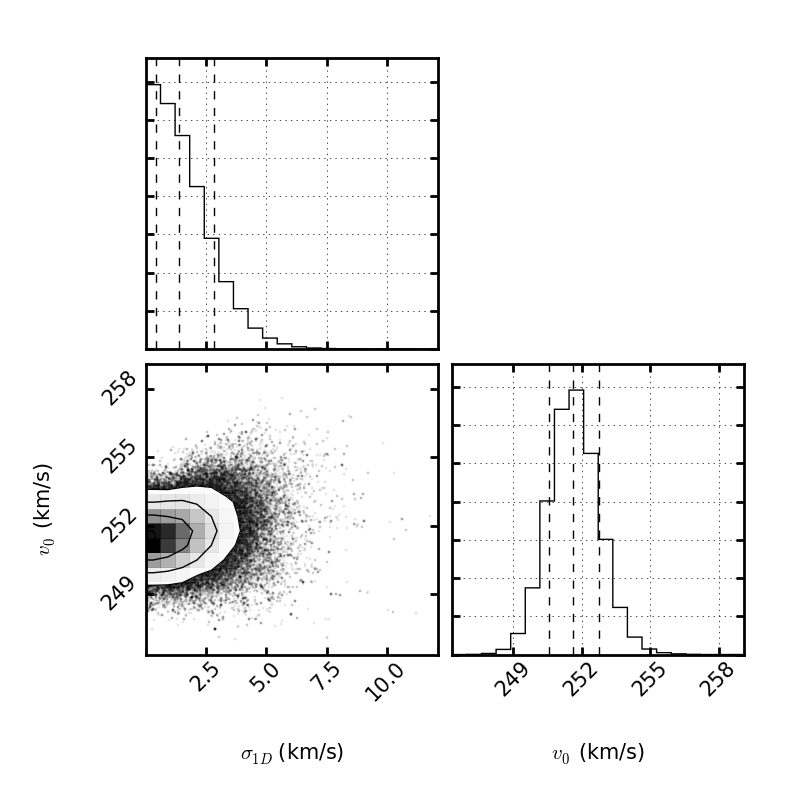}
 \caption{One- and two-dimensional projections of the posterior probability distributions of the line-of-sight velocity dispersion ($\sigma_{1D}$) and systemic velocity ($v_{0}$) for NGC\,2100 assuming the dispersion is Gaussian and constant over the range measured.
 Using this method the velocity of NGC\,2100 is 251.6\,$\pm$\,1.1\,\kms.
 This figure also demonstrates that the velocity dispersion for the sample is unresolved, we can therefore place an upper limit on $\sigma_{1D} <$~3.9\,\kms~at the 95\% confidence level.
 The vertical dashed lines in these figures indicate the estimated parameters with their associated 1\,$\sigma$ uncertainties.
\label{fig:sig1d}
          }
\end{figure}

\begin{figure}
 \includegraphics[width=9.0cm]{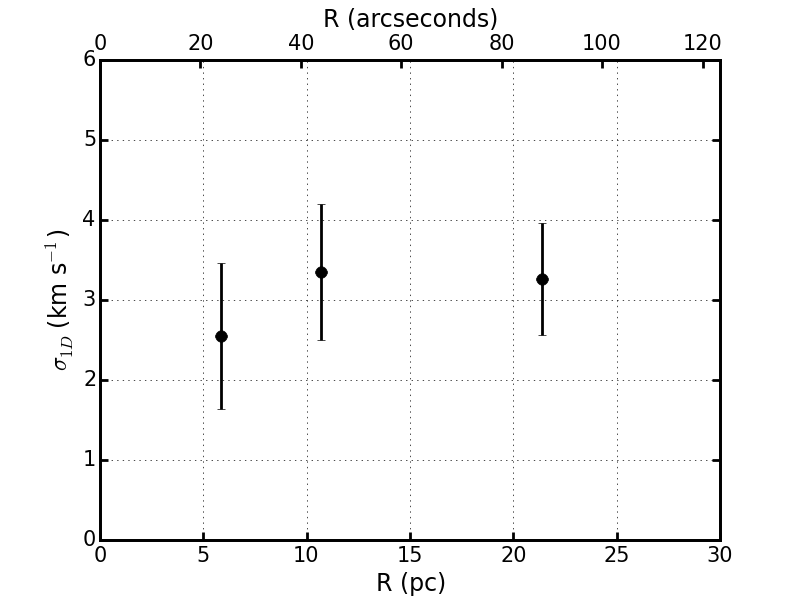}
 \caption{Upper limits to the line-of-sight velocity dispersion for the NGC\,2100 RSGs in three radial bins as a function of the distance from the centre of NGC\,2100.
 This figure demonstrates that we find no evidence for spatial variations in $\sigma_{1D}$.
 However, we note that in each radial bin the underlying dispersion is unresolved (see Figure~\ref{fig:sig1d}).
\label{fig:sig1dbins}
          }
\end{figure}

As shown in Figure~\ref{fig:sig1d}, the line-of-sight velocity dispersion ($\sigma_{1D}$) of NGC\,2100 is unresolved given the current data.
We can therefore place an upper limit on $\sigma_{1D} <$~3.9\,\kms~at the 95\% confidence level.
In Figure~\ref{fig:sig1dbins} we demonstrate that we find no evidence for spatial variations in the measured $\sigma_{1D}$ and we note that in each radial bin (which contain 5, 4 and 5 stars respectively), the measured dispersion is unresolved.

\subsection{Dynamical mass} 
\label{sub:dynamical_mass}
Using $\sigma_{1D}$ as an upper limit on the velocity distribution, one can calculate an upper limit on dynamical mass of the cluster using the virial equation:

\begin{equation}
  M_{dyn} = \frac{\eta\sigma_{1D}^{2}r_{\rm eff}}{G}
  \label{eq:vir}
\end{equation}

\noindent where $M_{dyn}$ is the dynamical mass and $\eta$~=~6$r_{vir}/r_{\rm eff}$~=~9.75 -- providing the density profile of the cluster is sufficiently steep~\citep{2010ARA&A..48..431P} --
where $r_{\rm eff}$~=~4.41\,pc for NGC\,2100~\citep{2005ApJS..161..304M}.
However, NGC\,2100 has a relatively shallow density profile~\citep[$\gamma$~=~$2.44\pm0.14$;][]{2003MNRAS.338...85M}
which means $\eta$~$<$~9.75.
Using $\sigma_{1D}$~=~$3.9$\,\kms~and equation~\ref{eq:vir}, an upper limit on the dynamical mass of NGC\,2100 is $M_{dyn}$~=~$15.2\times 10^{4}M_{\odot}$.
Comparing this to the photometric mass $M_{phot}$~=~$(2.3\pm1.0)\times 10^{4}M_{\odot}$~\citep{2005ApJS..161..304M},
we see that the upper limit on the dynamical mass is larger.

As discussed by~\citet{2010MNRAS.402.1750G}, binary motions can increase the measured velocity dispersion profile~\citep[e.g. see][]{2012A&A...546A..73H}.
However, as~\citet{2010MNRAS.402.1750G} note, the mean lifetime for RSGs in binary systems is significantly decreased and, where mass transfer occurs, their number decrease dramatically~\citep{2008MNRAS.384.1109E}.
We therefore expect that the number of RSGs in close binaries is small~\citep{1979MNRAS.186..831F,2009ApJ...696.2014D}.
The fraction of RSGs in longer-period systems is less certain, but these would contribute substantially less to the line-of-sight velocity distribution.

These arguments suggest that our estimate for the velocity dispersion in NGC\,2100 is not significantly increased by binary motions as our target stars are expected to be (predominantly) single objects. As the true dispersion of the cluster appears to be unresolved (Figure~\ref{fig:sig1d}), we conclude therefore that the upper limit of the dynamical mass is consistent with the published photometric mass.

Evidence in the literature suggests that J054211.61$-$691309.2 is an eclipsing binary system VV Cep~\citep{1979MNRAS.186..831F}.
The radial velocity of this star (256.6\,$\pm$\,6.1\,\kms) appears to be slightly enhanced with respect to systemic velocity of NGC\,2100, however, further study would be required to unambiguously classify this object as a binary.


\subsection{Stellar parameters} 
\label{sub:stellar_parameters}

Stellar parameters are estimated for each target using the $J$-band analysis technique described initially by~\cite{2010MNRAS.407.1203D}
and tested rigorously by~\cite{2014ApJ...788...58G} and~\cite{2015ApJ...806...21D}.
These studies show that by using a narrow spectral window within the $J$-band one can accurately derive overall metallicities
([Z]~=~$\log$(Z/Z$_{\odot}$)) to better than
$\pm$\,0.15\,dex at the resolution of KMOS observations with S/N~$\ge~100$.
\cite{2015ApJ...803...14P} built on this by demonstrating the feasibility of this technique using KMOS spectra.

The analysis uses synthetic RSG spectra, extracted from {\sc marcs} model atmospheres~\citep{2008A&A...486..951G},
computed with corrections for non-local thermodynamic equilibrium for lines from titanium, iron, silicon and magnesium
\citep{2012ApJ...751..156B,2013ApJ...764..115B,2015ApJ...804..113B}.
The parameter ranges for the grid of synthetic RSG spectra are listed in Table~\ref{tb:mod_range}.
The synthetic spectra are compared with observations using the $\chi$-squared statistic and the synthetic spectra are degraded to the resolution and sampling of the observations.
The diagnostic spectral features used to estimate stellar parameters have equal weighting in the analysis.

\begin{figure*}
 \begin{center}
\includegraphics[width=0.75\textwidth]{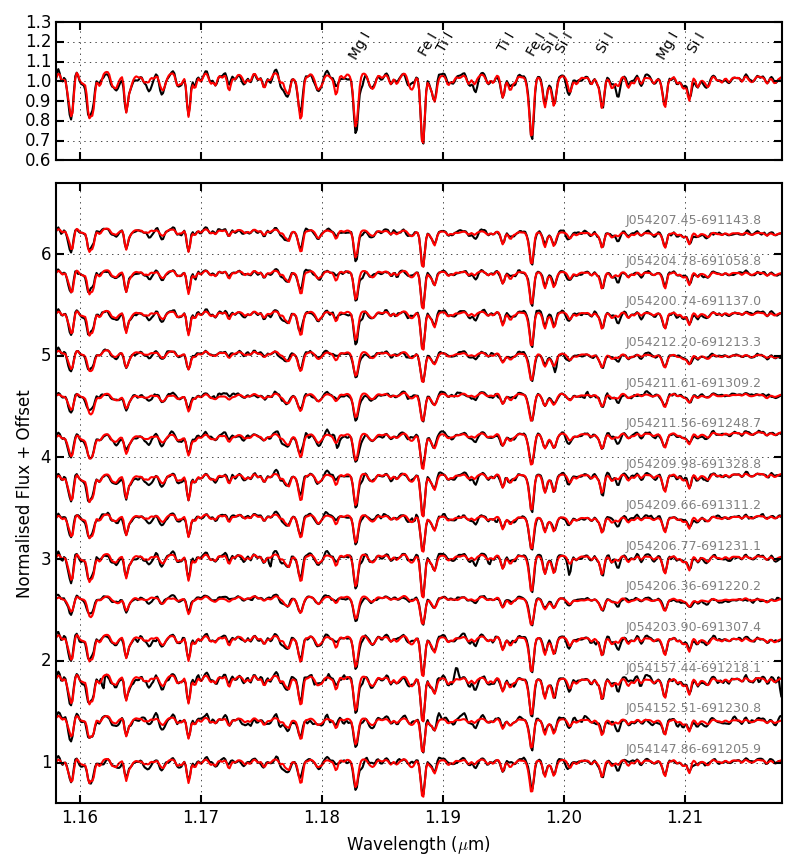}
\caption{KMOS spectra of RSGs in NGC\,2100 and their associated best-fit models
(black and red lines, respectively).
The upper panel shows the simulated integrated-light cluster spectrum;
the lower panel shows spectra for the individual RSGs.
The lines used for the analysis, from left-to-right by species, are
Fe\,{\scriptsize I}$\,\lambda\lambda$1.188285,
1.197305;
Mg\,{\scriptsize I}$\,\lambda\lambda$1.182819,
1.208335;
Si\,{\scriptsize I}$\,\lambda\lambda$1.198419,
1.199157,
1.203151,
1.210353;
Ti\,{\scriptsize I}$\,\lambda\lambda$1.189289,
1.194954.\label{fig:model_fits}}
\end{center}
\end{figure*}

Estimated stellar parameters are listed in Table~\ref{tb:stellar-params}.
Figure~\ref{fig:model_fits} shows the observed KMOS spectra (black) compared to their best-fitting models (red).
The average metallicity for the 14 RSGs is [Z]~=~$-$0.38\,$\pm$\,0.20\,dex where the large scatter is a result of the contribution from (J054211.61$-$691309.2).
Excluding this apparent outlier yields an average metallicity of [Z]~=~$-$0.43\,$\pm$\,0.10\,dex, which reduces the scatter and does not alter the result significantly.
The model fit parameters of J054211.61$-$691309.2 suggest a considerably ($\times$1.7) super-solar
metallicity.
This appears unlikely given its apparent membership to the LMC, and it is notable that the estimates for the surface gravity and microturbulence parameters are also outliers compared to the rest of the sample.
In addition, as noted above, this star was flagged as a potential eclipsing binary by~\citep{1979MNRAS.186..831F}, therefore this target is excluded from the sample in further analysis.

The average metallicity in NGC\,2100 estimated here is in good agreement with estimates of the cluster metallicity using isochrone fitting to the optical colour-magnitude diagram~\citep[$-$0.34\,dex;][]{2015A&A...575A..62N}.
The only other estimate of stellar metallicity within this cluster is from JT94
who estimated metallicities using optical spectroscopy of four RSGs.
These authors found an average metallicity for NGC\,2100 of [Fe/H]~=~$-$0.32\,$\pm$\,0.03\,dex, which is in reasonable agreement with our estimate.
There are three targets in common with our study: B17, C2 and C32
\citep[using the][nomenclature]{1974A&AS...15..261R}.
Given the differences in the analyses (i.e. optical cf. infrared, and the different models used) the estimated parameters are in reasonable agreement for all three stars
(aside from the spectroscopic gravities quoted by JT94, but with reasonable agreement with their photometric gravity estimates).

\begin{figure}
 \includegraphics[width=9.0cm]{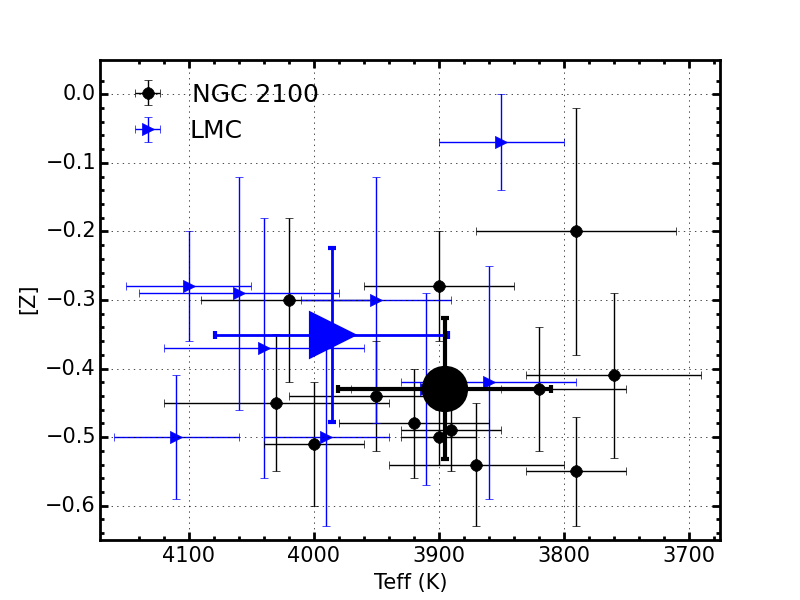}
 \caption{Estimated metallicities for NGC\,2100 RSGs in this study shown against effective temperature (black points).
        For comparison we show the distribution of LMC RSGs from~\citet[][blue triangles]{2015ApJ...806...21D} with good agreement between the means of the two samples.
\label{fig:TeffvsZ}
          }
\end{figure}

Using the same analysis technique as in this study,
\cite{2015ApJ...806...21D} estimate metallicities for nine RSGs within the LMC,
finding an average value of [Z]~=~$-$0.37\,$\pm$\,0.14\,dex, which our estimate agrees well with.
In Figure~\ref{fig:TeffvsZ}, we compare the effective temperatures and metallicities from NGC\,2100 with those estimated for RSGs elsewhere in the LMC.
We find good agreement in the distribution of temperatures from the two studies, with the average agreeing well.
The range in [Z] from the LMC population is slightly larger than that of the NGC\,2100 RSGs, which is expected when comparing a star cluster with an entire galaxy; however, the averages for the two studies agree very well.

\begin{table}
\caption{
Model grid used for the spectroscopic analysis.\label{tb:mod_range}
         }
\scriptsize
\begin{center}
\begin{tabular}{lccc}
 \hline
 \hline
  Model Parameter & Min. & Max. & Step size \\
 \hline
T$_{\rm eff}$ (K)       & 3400 & 4400 & 100 \\
$[$Z$]$ (dex)   & $-$1.0\o & 1.0\o  & 0.1\o\\
log\,$g$ (cgs)  & $-$1.00 & 1.00 & 0.25\\
 $\xi$ (\kms)  & \pp1.0\o & 5.0\o & 0.2\o\\
 \hline
\end{tabular}
\end{center}
\end{table}

\begin{table*}
\begin{center}
\caption{
Physical parameters determined for the KMOS targets in NGC\,2100.
\label{tb:stellar-params}
         }
\scriptsize
\begin{threeparttable}
\begin{tabular}{lc ccccl}
 \hline
 \hline
  Target  & IFU & $\xi$ (\kms) & [Z] & log\,$g$ & T$_{\rm eff}$ (K) & Notes\tnote{a}\\
  \hline
J054147.86$-$691205.9 & 7  & 3.6\,$\pm$\,0.2 & $-$0.45\,$\pm$\,0.10 & 0.10\,$\pm$\,0.16 & 4030\,$\pm$\,90\o & D15\\
J054152.51$-$691230.8 & 9  & 3.6\,$\pm$\,0.2 & $-$0.51\,$\pm$\,0.09 & 0.43\,$\pm$\,0.18 & 4000\,$\pm$\,40\o & D16\\
J054157.44$-$691218.1 & 6  & 4.9\,$\pm$\,0.1 & $-$0.44\,$\pm$\,0.08 & 0.15\,$\pm$\,0.20 & 3950\,$\pm$\,70\o & C2\\ 
J054200.74$-$691137.0 & 4  & 4.2\,$\pm$\,0.2 & $-$0.55\,$\pm$\,0.08 & 0.23\,$\pm$\,0.10 & 3790\,$\pm$\,40\o & C8\\
J054203.90$-$691307.4 & 12 & 4.5\,$\pm$\,0.2 & $-$0.49\,$\pm$\,0.06 & 0.23\,$\pm$\,0.09 & 3890\,$\pm$\,40\o & B4\\
J054204.78$-$691058.8 & 3  & 4.2\,$\pm$\,0.2 & $-$0.54\,$\pm$\,0.09 & 0.46\,$\pm$\,0.15 & 3870\,$\pm$\,70\o & \ldots\\
J054206.36$-$691220.2 & 24 & 2.8\,$\pm$\,0.4 & $-$0.20\,$\pm$\,0.18 & 0.42\,$\pm$\,0.18 & 3790\,$\pm$\,80\o & B17\\
J054206.77$-$691231.1 & 10 & 4.9\,$\pm$\,0.2 & $-$0.50\,$\pm$\,0.04 & 0.25\,$\pm$\,0.09 & 3900\,$\pm$\,30\o & A127\\
J054207.45$-$691143.8 & 2  & 4.0\,$\pm$\,0.2 & $-$0.43\,$\pm$\,0.09 & 0.45\,$\pm$\,0.17 & 3820\,$\pm$\,70\o & C12\\
J054209.66$-$691311.2 & 14 & 3.8\,$\pm$\,0.2 & $-$0.41\,$\pm$\,0.12 & 0.06\,$\pm$\,0.20 & 3760\,$\pm$\,70\o & B47\\
J054209.98$-$691328.8 & 11 & 4.8\,$\pm$\,0.1 & $-$0.48\,$\pm$\,0.08 & 0.17\,$\pm$\,0.22 & 3920\,$\pm$\,60\o & C32\\
J054211.56$-$691248.7 & 20 & 3.8\,$\pm$\,0.2 & $-$0.28\,$\pm$\,0.08 & 0.01\,$\pm$\,0.16 & 3900\,$\pm$\,60\o & B40\\
J054211.61$-$691309.2 & 18 & 2.2\,$\pm$\,0.4 & \pp0.23\,$\pm$\,0.23 & 0.65\,$\pm$\,0.19 & 3800\,$\pm$\,100 & B46\\
J054212.20$-$691213.3 & 22 & 3.3\,$\pm$\,0.2 & $-$0.30\,$\pm$\,0.12 & 0.33\,$\pm$\,0.31 & 4020\,$\pm$\,70\o & B22\\
\\
NGC\,2100 average\tnote{b} & & 4.0\,$\pm$\,0.6 & $-$0.43\,$\pm$\,0.10 & 0.25\,$\pm$\,0.15 & 3890\,$\pm$\,85\o\\
\\
Integrated-light spectrum\tnote{c}             & & 4.6\,$\pm$\,0.3 & $-$0.42\,$\pm$\,0.14 & 0.37\,$\pm$\,0.22 & 3860\,$\pm$\,85\o\\
  \hline
  \end{tabular}
\begin{tablenotes}
    \item [a] ID in final column from{~\cite{1974A&AS...15..261R}}.
    \item [b] Averages computed excluding J054211.61$-$691309.2. See text for details.
    \item [c] Simulated integrated light cluster spectrum parameters estimated excluding J054211.61$-$691309.2.
\end{tablenotes}
  \end{threeparttable}
  \end{center}
\end{table*}


\section{Discussion} 
\label{sec:discussion}

\subsection{Stellar parameters} 
\label{sub:stellar_parameters_disc}

\begin{figure}
 \includegraphics[width=9.0cm]{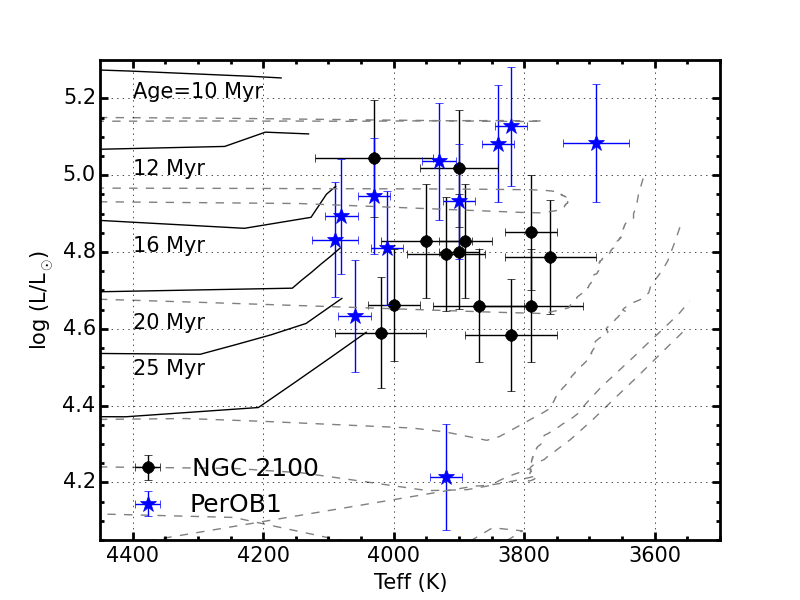}
 \caption{H--R diagram for 14 RSGs in NGC\,2100 (black points).
  Isochrones for solar~\citep[dashed grey lines;][]{2012A&A...537A.146E} and SMC-like~\citep[solid black lines;][]{2013A&A...558A.103G} metal abundances,
  in which stellar rotation is 40\% of the break-up velocity, are shown for ages of 10-32\,Myr. For comparison, 11 RSGs from the Galactic YMC Perseus OB-1 are overlaid with blue stars~\citep{2014ApJ...788...58G}.
  The best-fit isochrone to the observed data has an age of 20\,$\pm$\,5\,Myr for both SMC- and solar-like metallicities.
  \label{fig:HRD}
          }
\end{figure}

Luminosities have been estimated for our sample from {\it K}-band photometry (see Table~\ref{tb:obs-params}) using the bolometric correction from~\cite{2013ApJ...767....3D} with a small contribution from interstellar extinction using E(B$-$V)~=~0.17~\citep{2015A&A...575A..62N} assuming $R_V$~=~3.5~\citep{2013A&A...558A.134D} and $A_K/A_V$~=~0.112~\citep{1985ApJ...288..618R}.
The H--R diagram for the cluster is presented in Figure~\ref{fig:HRD}.
Overlaid on this H--R diagram are {\sc syclist} stellar isochrones for SMC-like~\citep[solid lines;][]{2013A&A...558A.103G} and Solar-like~\citep[dashed lines;][]{2012A&A...537A.146E} models, where stellar rotation is 40\% of break-up velocity.
Even though the temperatures covered by the SMC-like models do not represent the distribution of temperatures observed in this study, they remain useful to constrain the age of NGC\,2100.
The Solar-like models (dashed) demonstrate that, when compared with the SMC-like models, increasing the metallicity of the sample
(a) decreases the average temperature of the RSGs~\citep[something which is not observed by][]{2015ApJ...803...14P},
(b) induces so-called `blue loop' behaviour for the youngest models and
(c) decreases the luminosity for the youngest models.
The best-fitting model to the observed data has an age of 20\,$\pm$\,5\,Myr, in reasonable agreement with the estimate in Beasor \& Davies (submitted).

In addition, results for 11 RSGs from the Galactic star cluster Perseus OB-1~\citep[PerOB1;][]{2014ApJ...787..142G} are overlaid in Figure~\ref{fig:HRD} (blue stars) for which stellar parameters were estimated using the same analysis technique as in this study.
PerOB1 is a cluster with a similar mass and age~\citep[$2\times10^{4}\,$M$_{\odot}$ and 14\,Myr respectively;][]{2010ApJS..186..191C}
as NGC\,2100, and a comparison between the stellar components of these two clusters using a consistent analysis technique is useful to highlight differences in stellar evolution within clusters at this range of metallicities.

We can see from Figure~\ref{fig:HRD} that, generally, the estimated temperatures are in good agreement between the two clusters.
The median luminosity for the PerOB1 targets ($10^{4.93\pm0.15}\,$L$_{\odot}$) is slightly above that of NGC\,2100 ($10^{4.77\pm0.15}\,$L$_{\odot}$) which could represent the slight difference in the ages of the two clusters.
As PerOB1 is younger, the average mass for a RSG in the cluster will be larger than the average in NGC\,2100.
Therefore, we would expect to see higher luminosity RSGs in PerOB1.
However, the difference between the two samples is barely significant and is consistent with a constant luminosity.
The average effective temperatures for the two data sets (NGC\,2100: 3890\,$\pm$\,20\,K, PerOB1: 3940\,$\pm$\,10\,K) are in reasonable agreement, where the spread in temperatures is slightly larger for PerOB1
($\sigma_{\rm PerOB1}$~=~120\,K, $\sigma_{\rm NGC\,2100}$~=~85\,K),
particularly so for the highest luminosity targets within the PerOB1 sample.
Overall, by comparing these two star clusters with a similar mass, age and stellar population, we conclude that there exists no significant difference in appearance on the H--R diagram of RSGs within these star clusters of different metallicities.

The stellar parameters have been estimated assuming a Solar-like [$\alpha$/Fe]~=~0.0.
As we have recently included the non-LTE corrections for two strong Mg\,\1 lines~\citep{2015ApJ...804..113B} we now have the tools to estimate the [$\alpha$/Fe] given that we have increased the number of diagnostic lines used as well as the number of $\alpha$ elements.
We do not include [$\alpha$/Fe] as a free parameter in our model as the [$\alpha$/Fe] for the LMC appears to be within $\pm$\,0.2~\citep[see][and references therein]{2015ApJ...806...21D}.
In the near future we will introduce the [$\alpha$/Fe] as a free parameter in this analysis routine where its dependencies and any potential degeneracies will be quantified rigorously, however, this is beyond the scope of the present study.


\subsection{Simulated cluster spectrum analysis} 
\label{sub:integrated_light_cluster_analysis}

We can use the individual stars in NGC\,2100 to simulate the analysis of a YMC in the more distant Universe, using the assumption that RSGs dominate the near-IR flux from such a cluster~\citep{2013MNRAS.430L..35G}.
\cite{2014ApJ...788...58G} use this assumption to create a simulated integrated-light cluster spectrum for PerOB1 and show that, by analysing the combined spectrum from their 11 RSGs, the resulting parameters are consistent with the average parameters estimated using the individual stars.
NGC\,2100 has a similar mass and age to PerOB1 and~\cite{2014ApJ...788...58G} study a similar number of RSGs to this study,
therefore, a direct comparison between the two clusters is useful to investigate potential metallicity dependencies.

To create a simulated integrated-light cluster spectrum we sum all the individual RSG spectra weighted by their $J$-band luminosities.
The resulting spectrum is then degraded to the lowest resolution spectrum of the sample using a simple Gaussian filter.
The top panel of Figure~\ref{fig:model_fits} shows the resulting integrated-light cluster spectrum.
This spectrum is then analysed in the same way described in Section~\ref{sub:stellar_parameters} for a single RSG.
The results of this analysis are what one would expect from KMOS observations of more distant YMCs where individual stars cannot be resolved.
We find a metallicity of $-$0.42\,$\pm$\,0.14\,dex, an effective temperature of 3860\,$\pm$\,85\,K,
a surface gravity of 0.37\,$\pm$\,0.22\,dex and a microturbulent velocity of 4.6\,$\pm$\,0.3\,\kms~which agree well with the averages of the individual RSG parameters.


\subsection{Velocity dispersion and dynamical mass} 
\label{sub:velocity_dispersion_Mdyn}

This study represents the first estimate of an upper limit to the line-of-sight velocity dispersion profile for NGC\,2100.
Comparing this estimate with that of other YMCs in the Local Universe is useful to ascertain if this cluster shares similar properties with other YMCs.
We find the properties NGC\,2100 are well matched by other clusters with similar masses and ages, particularly so with RSGC01, a Galactic YMC~\citep{2007ApJ...671..781D}.

Owing to the non-negligible contribution from measurement errors, the $\sigma_{1D}$ adopted here is an upper limit to the true dispersion within the cluster which is likely to be significantly smaller.
Using the data available, we can rule out an $\sigma_{1D}$ value significantly larger than 3.9\,\kms, however,
the true dispersion of the cluster is unresolved.

By extension, the dynamical mass estimated here is therefore also an upper limit to the true mass of the cluster.
There are several factors that could alter the value of the dynamical mass estimate.
The likely value of the $\eta$ parameter is discussed in Section~\ref{sub:dynamical_mass} and any change in this value will act to decrease the estimated dynamical mass.


\section{Conclusions} 
\label{sec:conclusions}

Using KMOS spectra of 14 RSGs in NGC\,2100 we have estimated the dynamical properties of this YMC for the first time.
Radial velocities have been estimated to a precision of $<~5$\,\kms demonstrating that KMOS can be used to study the dynamical properties of star clusters in external galaxies.

An upper limit to the average line-of-sight velocity dispersion of $\sigma_{1D}$~=~3.9\,\kms~has been estimated, at the 95\% confidence level, and we find no evidence for spatial variations.
Using the average velocity dispersion within NGC\,2100 allows an upper limit on the dynamical mass to be calculated
(assuming virial equilibrium) as $M_{dyn}$~=~$15.2\times 10^{4}M_{\odot}$.
This measurement is consistent with the literature measurement of the photometric mass~\citep{2005ApJS..161..304M} as the true dispersion is unresolved.

In addition to estimating the dynamical properties of NGC\,2100, we have also estimated stellar parameters for the RSGs in NGC\,2100 using the new $J$-band analysis technique~\citep{2010MNRAS.407.1203D}.
We find the average metallicity for RSGs in NGC\,2100 is [Z]~=~$-$0.43\,$\pm$\,0.10\,dex, which agrees well with previous studies within this cluster and with studies of the young stellar population of the LMC.

The H--R diagram of NGC\,2100 is compared with that of PerOB1: a Galactic YMC with a similar age, mass and stellar population.
Using stellar parameters estimated from~\cite{2014ApJ...788...58G}, obtained with the same technique as in this study, we demonstrate that there is no significant difference in the appearance of the H--R diagram of YMCs between Solar- and LMC-like metallicities.

By combining the individual RSG spectra within NGC\,2100, we have simulated an integrated-light cluster spectrum and proceeded to analyse this spectrum using the same techniques for that of the individual RSGs, as RSGs dominate the cluster light in the $J$-band~\citep{2013MNRAS.430L..35G}.
The results of this technique demonstrate the potential of this analysis for integrated light spectra of more distant YMCs in low-metallicity environments.
We find good agreement using the integrated-light cluster spectrum with the average results of the individual RSGs.

\section*{Acknowledgements}
The authors would like to thank the anonymous referee for a careful review which has improved the quality of this publication.
In addition, we would like to thank A. Hall for discussions surrounding Bayesian probabilities, M.~Gieles and A.-L. Varri for helpful discussions and suggestions.
Based on observations collected at the European Organisation for Astronomical Research in the Southern Hemisphere under ESO programme 095.B-0022(A).


\begin{thebibliography}{99}




\bibitem[Bergemann et al.(2012)]{2012ApJ...751..156B} Bergemann, M.,
Kudritzki, R.-P., Plez, B., et al.\ 2012, \apj, 751, 156

\bibitem[Bergemann et al.(2013)]{2013ApJ...764..115B} Bergemann, M.,
Kudritzki, R.-P., W{\"u}rl, M., et al.\ 2013, \apj, 764, 115

\bibitem[Bergemann et al.(2015)]{2015ApJ...804..113B} Bergemann, M.,
Kudritzki, R.-P., Gazak, Z., Davies, B., \& Plez, B.\ 2015, \apj, 804, 113

\bibitem[Cabrera-Ziri et al.(2014)]{2014MNRAS.441.2754C} Cabrera-Ziri, I.,
Bastian, N., Davies, B., et al.\ 2014, \mnras, 441, 2754

\bibitem[Cioni et
al.(2011)]{2011A&A...527A.116C} Cioni, M.-R.~L., Clementini, G., Girardi, L., et al.\ 2011, \aap, 527, A116


\bibitem[Crowther et al.(2010)]{2010MNRAS.408..731C} Crowther, P.~A.,
Schnurr, O., Hirschi, R., et al.\ 2010, \mnras, 408, 731

\bibitem[Currie et al.(2010)]{2010ApJS..186..191C} Currie, T., Hernandez,
J., Irwin, J., et al.\ 2010, \apjs, 186, 191

\bibitem[Davies et al.(2007)]{2007ApJ...671..781D} Davies, B., Figer,
D.~F., Kudritzki, R.-P., et al.\ 2007, \apj, 671, 781

\bibitem[Davies et al.(2009)]{2009ApJ...696.2014D} Davies, B., Origlia, L.,
Kudritzki, R.-P., et al.\ 2009, \apj, 696, 2014

\bibitem[Davies et al.(2010)]{2010MNRAS.407.1203D} Davies, B., Kudritzki,
R.-P., \& Figer, D.~F.\ 2010, \mnras, 407, 1203

\bibitem[Davies, B., et al.(2013)]{2013ApJ...767....3D} Davies, B., Kudritzki,
R.-P., Plez, B., et al.\ 2013, \apj, 767, 3

\bibitem[Davies et al.(2015)]{2015ApJ...806...21D} Davies, B., Kudritzki,
R.-P., Gazak, Z., et al.\ 2015, \apj, 806, 21

\bibitem[Davies, R.~I., et
al.(2013)]{2013A&A...558A..56D} Davies, R.~I., Agudo Berbel, A., Wiezorrek, E., et al.\ 2013, \aap, 558, A56


\bibitem[de Grijs et al.(2014)]{2014AJ....147..122D} de Grijs, R., Wicker,
J.~E., \& Bono, G.\ 2014, \aj, 147, 122

\bibitem[de Wit et
al.(2005)]{2005A&A...437..247D} de Wit, W.~J., Testi, L., Palla, F., \& Zinnecker, H.\ 2005, \aap, 437, 247

\bibitem[Doran et
al.(2013)]{2013A&A...558A.134D} Doran, E.~I., Crowther, P.~A., de Koter, A., et al.\ 2013, \aap, 558, A134

\bibitem[Ekstr{\"o}m et
al.(2012)]{2012A&A...537A.146E} Ekstr{\"o}m, S., Georgy, C., Eggenberger, P., et al.\ 2012, \aap, 537, A146

\bibitem[Eldridge et al.(2008)]{2008MNRAS.384.1109E} Eldridge, J.~J.,
Izzard, R.~G., \& Tout, C.~A.\ 2008, \mnras, 384, 1109

\bibitem[Elson(1991)]{1991ApJS...76..185E} Elson, R.~A.~W.\ 1991, \apjs,
76, 185


\bibitem[Evans et
al.(2011)]{2011A&A...527A..50E} Evans, C.~J., Davies, B., Kudritzki, R.-P., et al.\ 2011, \aap, 527, A50


\bibitem[Evans et
al.(2015)]{2015A&A...584A...5E} Evans, C.~J., van Loon, J.~T., Hainich, R., \& Bailey, M.\ 2015, \aap, 584, A5

\bibitem[Feast(1979)]{1979MNRAS.186..831F} Feast, M.~W.\ 1979, \mnras, 186,
831



\bibitem[Foreman-Mackey et al.(2013)]{2013PASP..125..306F} Foreman-Mackey,
D., Hogg, D.~W., Lang, D., \& Goodman, J.\ 2013, \pasp, 125, 306

\bibitem[Gazak et al.(2013)]{2013MNRAS.430L..35G} Gazak, J.~Z., Bastian,
N., Kudritzki, R.-P., et al.\ 2013, \mnras, 430, L35

\bibitem[Gazak et al.(2014a)]{2014ApJ...788...58G} Gazak, J.~Z., Davies, B.,
Kudritzki, R., Bergemann, M., \& Plez, B.\ 2014a, \apj, 788, 58

\bibitem[Gazak et al.(2014b)]{2014ApJ...787..142G} Gazak, J.~Z., Davies, B.,
Bastian, N., et al.\ 2014b, \apj, 787, 142

\bibitem[Gazak et al.(2015)]{2015ApJ...805..182G} Gazak, J.~Z., Kudritzki,
R., Evans, C., et al.\ 2015, \apj, 805, 182

\bibitem[Georgy et
al.(2013)]{2013A&A...558A.103G} Georgy, C., Ekstr{\"o}m, S., Eggenberger, P., et al.\ 2013, \aap, 558, A103

\bibitem[Gieles et al.(2010)]{2010MNRAS.402.1750G} Gieles, M., Sana, H.,
\& Portegies Zwart, S.~F.\ 2010, \mnras, 402, 1750



\bibitem[Goodman \& Weare(2010)]{2010CAMCS.5..65G} Goodman, J., \&
Weare, J.\ 2010, Comm. App. Math. Comp. Sci., 5(1), 65

\bibitem[Gratton et
al.(2012)]{2012A&ARv..20...50G} Gratton, R.~G., Carretta, E., \& Bragaglia, A.\ 2012, \aapr, 20, 50


\bibitem[Gustafsson et
al.(2008)]{2008A&A...486..951G} Gustafsson, B., Edvardsson, B., Eriksson, K., et al.\ 2008, \aap, 486, 951

\bibitem[H{\'e}nault-Brunet et
al.(2012)]{2012A&A...546A..73H} H{\'e}nault-Brunet, V., Evans, C.~J., Sana, H., et al.\ 2012, \aap, 546, A73

\bibitem[Jasniewicz
\& Thevenin(1994)]{1994A&A...282..717J} Jasniewicz, G., \& Thevenin, F.\ 1994, \aap, 282, 717

\bibitem[King(1966)]{1966AJ.....71...64K} King, I.~R.\ 1966, \aj, 71, 64

\bibitem[Lada
\& Lada(2003)]{2003ARA&A..41...57L} Lada, C.~J., \& Lada, E.~A.\ 2003, \araa, 41, 57

\bibitem[Lapenna et al.(2015)]{2015ApJ...798...23L} Lapenna, E., Origlia,
L., Mucciarelli, A., et al.\ 2015, \apj, 798, 23

\bibitem[Lardo et al.(2015)]{2015ApJ...812..160L} Lardo, C., Davies, B.,
Kudritzki, R.-P., et al.\ 2015, \apj, 812, 160


\bibitem[Mackey
\& Gilmore(2003)]{2003MNRAS.338...85M} Mackey, A.~D., \& Gilmore, G.~F.\ 2003, \mnras, 338, 85

\bibitem[Massey
\& Hunter(1998)]{1998ApJ...493..180M} Massey, P., \& Hunter, D.~A.\ 1998, \apj, 493, 180


\bibitem[McLaughlin
\& van der Marel(2005)]{2005ApJS..161..304M} McLaughlin, D.~E., \& van der Marel, R.~P.\ 2005, \apjs, 161, 304

\bibitem[Miller et al.(1997)]{1997AJ....114.2381M} Miller, B.~W., Whitmore,
B.~C., Schweizer, F., \& Fall, S.~M.\ 1997, \aj, 114, 2381

\bibitem[Niederhofer et
al.(2015)]{2015A&A...575A..62N} Niederhofer, F., Hilker, M., Bastian, N., \& Silva-Villa, E.\ 2015, \aap, 575, A62

\bibitem[Parker
\& Goodwin(2007)]{2007MNRAS.380.1271P} Parker, R.~J., \& Goodwin, S.~P.\ 2007, \mnras, 380, 1271

\bibitem[Patrick et al.(2015)]{2015ApJ...803...14P} Patrick, L.~R., Evans,
C.~J., Davies, B., et al.\ 2015, \apj, 803, 14

\bibitem[Pietrzy{\'n}ski et al.(2013)]{2013Natur.495...76P}
Pietrzy{\'n}ski, G., Graczyk, D., Gieren, W., et al.\ 2013, \nat, 495, 76

\bibitem[Points et al.(1999)]{1999ApJ...518..298P} Points, S.~D., Chu,
Y.~H., Kim, S., et al.\ 1999, \apj, 518, 298

\bibitem[Portegies Zwart et
al.(2010)]{2010ARA&A..48..431P} Portegies Zwart, S.~F., McMillan, S.~L.~W., \& Gieles, M.\ 2010, \araa, 48, 431

\bibitem[Rieke
\& Lebofsky(1985)]{1985ApJ...288..618R} Rieke, G.~H., \& Lebofsky, M.~J.\ 1985, \apj, 288, 618


\bibitem[Robertson(1974)]{1974A&AS...15..261R} Robertson, J.~W.\ 1974, \aaps, 15, 261


\bibitem[Sharples et al.(2013)]{2013Msngr.151...21S} Sharples, R., Bender,
R., Agudo Berbel, A., et al.\ 2013, The Messenger, 151, 21



\bibitem[Whitmore
\& Schweizer(1995)]{1995AJ....109..960W} Whitmore, B.~C., \& Schweizer, F.\ 1995, \aj, 109, 960

\bibitem[Zepf et al.(1999)]{1999AJ....118..752Z} Zepf, S.~E., Ashman,
K.~M., English, J., Freeman, K.~C., \& Sharples, R.~M.\ 1999, \aj, 118, 752
\end{thebibliography}

\label{lastpage}

\end{document}